\documentclass[sigconf]{acmart}
\renewcommand\footnotetextcopyrightpermission[1]{} 

\usepackage{amsmath,amssymb,amsfonts}
\usepackage{algorithm}
\usepackage{algorithmic}
\usepackage{graphicx}
\usepackage{textcomp}
\usepackage{xcolor} 
\usepackage{subfigure}

\usepackage{enumitem}

\newtheorem{proposition}{Proposition}
\newtheorem{lemma}{Lemma}
\newtheorem{assumption}{Assumption}

\newlength\myindent
\newcommand\bindent{%
  \begingroup
  \setlength{\itemindent}{\myindent}
  \addtolength{\algorithmicindent}{\myindent}
}
\newcommand\eindent{\endgroup}

\begin{document}
\title{FDML: A Collaborative Machine Learning Framework for Distributed Features}

\author{Yaochen Hu$^1$, Di Niu$^1$, Jianming Yang$^2$, Shengping Zhou$^2$}
\affiliation{$^1$University of Alberta, Edmonton, AB, Canada}
\affiliation{$^2$Platform and Content Group, Tencent, Shenzhen, China}




\renewcommand{\shortauthors}{Y. Hu et al.}

\begin{abstract}
Most current distributed machine learning systems try to scale up model training by using a data-parallel architecture that divides the computation of gradients for different samples among workers. We study distributed machine learning from a different motivation, where the information about the same samples, e.g., users and objects, are owned by several parities that wish to collaborate but do not want to share raw data with each other. 
We propose an asynchronous stochastic gradient descent (SGD) algorithm for such a feature distributed machine learning (FDML) problem, to jointly learn from distributed features, with theoretical convergence guarantees under bounded asynchrony. Our algorithm does not require sharing the original features or even local model parameters between parties, thus preserving the data locality. The system can also easily incorporate differential privacy mechanisms to preserve a higher level of privacy. We implement the FDML system in a parameter server architecture and compare our system with fully centralized learning (which violates data locality) and learning based on only local features, through extensive experiments performed on both a public data set {\emph a9a}, and a large dataset of $5,000,000$ records and $8700$ decentralized features from three collaborating apps at Tencent including {\emph Tencent MyApp}, {\emph Tecent QQ Browser} and {\emph Tencent Mobile Safeguard}. Experimental results have demonstrated that the proposed FDML system can be used to significantly enhance app recommendation in  Tencent MyApp by leveraging user and item features from other apps, while preserving the locality and privacy of features in each individual app to a high degree.
\end{abstract}


\settopmatter{printacmref=false}
\fancyhead{}

%
%




\maketitle

\section{Introduction}

While the success of modern machine learning lays the foundation of many intelligent services, the performance of a model often depends on the availability of data. 
In most applications, however, a large quantity of useful data may be generated on and held by multiple parties. Collecting such data to a central site for training incurs extra management and business compliance overhead, privacy concerns, or even regulation and judicial issues. 
As a result, a number of distributed machine learning techniques have been proposed to collaboratively train a model by letting each party perform local model updates and exchange locally computed gradients  \cite{shokri2015privacy} or model parameters \cite{mcmahan2016communication} with the central server to iteratively improve model accuracy.
Most of the existing schemes, however, fall into the range of \emph{data parallel} computation, where the training samples are located on different parties. For example, different users hold different images to jointly train a classifier. Different organizations may contribute their individual corpora to learn a joint language model.

We study distributed machine learning based on another motivation, where different {\it features} of a same sample are held by different parties. 
The question is---can we improve the predictive power at one party by leveraging additional features from another domain or party, yet without requiring any party to share its features?
This is a real problem we are solving at \emph{{Tencent MyApp}}, one of the largest Android app stores in China, with a market share of 24.7\% in China in 2017. \emph{Tencent MyApp} performs app recommendation and activation rate prediction based on the user and app features logged in its own platform. However, it turns out other Tencent apps including  \emph{Tencent QQ Browser} and \emph{Tencent Mobile Safeguard} share a large number of common users with MyApp.
Since these apps may have complementary information about a user, such cross-domain knowledge from another app, if utilized, may help to train a joint model that can improve app recommendation and customer behavior preference prediction in MyApp. However, due to privacy and customer protection regulations, raw customer data are not to be shared across apps that belong to different departments. 


A natural question is---how can we train a joint machine learning model if the features of each training sample are located on multiple distributed parties? To make the solution practical with the most conservative assumption on information sharing, we bear the following goals:
\begin{itemize}
	\item 
	 To minimize information leakage, no party should share its feature set. Neither should any of its local model parameters be communicated to other parties.
	 \item The prediction made by the joint model should outperform the prediction made by each isolated model trained only with a single party's feature set, provided that such improvement from joint features also exists in centralized training.
	\item The joint model produced should approach the model trained in a centralized manner if all the features were collected centrally.
	\item The system should be efficient in the presence of both large numbers of features and samples.
\end{itemize}

To solve the above challenges, in this paper, we design, implement and extensively evaluate a practical 
Feature Distributed Machine Learning (FDML) system based on real-world datasets. For any supervised learning task, e.g., classification, our system enables each party to use an arbitrary model (e.g., logistic regression, factorization machine, SVM, and deep neural networks) to map its local feature set to a local prediction, while different local predictions are aggregated into a final prediction for classification via a ``hyper-linear structure,'' which is similar to softmax. The entire model is trained end-to-end using a mini-batched stochastic gradient descent (SGD) algorithm performed in the sense of stale synchronous parallel (SSP) \cite{ho2013more}, i.e., different parties are allowed to be at different iterations of parameter updates up to a bounded delay.

A highlight of our system is that during each training iteration, every party is solely responsible for updating its own local model parameters (local net) using its own mini-batch of local feature sets, and for each record, only needs to share its local prediction to the central server (or to other parties directly in a fully decentralized scenario). Since neither the original features nor the local model parameters of a party are transferred to any external sites, the FDML system preserves data locality and is much less vulnerable to model inversion attacks \cite{hitaj2017deep} targeting other collaborative learning algorithms \cite{shokri2015privacy, zhou2016convergence} that share model parameters between parties. Moreover, we further enhance the data privacy by adopting a differential-privacy-based method \cite{dwork2008differential, dwork2014algorithmic, shokri2015privacy}. by adding some perturbations to the shared local predictions.
 
We theoretically establish a convergence rate of $O(\frac{1}{\sqrt{T}})$ for the proposed asynchronous FDML algorithm under certain assumptions (including the bounded delay assumption \cite{ho2013more}), where $T$ is the number of iterations on (the slowest) party, which matches the standard convergence rate of fully centralized synchronous SGD training with a convex loss as well as that known for asynchronously distributed data-parallel SGD in SSP \cite{ho2013more}.

We developed a distributed implementation of FDML in a parameter server architecture, and conducted experiments based on both a public data set {\emph a9a} \cite{Dua:2017}, and a large dataset of $5,000,000$ samples and $8700$ decentralized features collected from three popular Tencent Apps, including \emph{Tencent MyApp}, \emph{Tencent QQ Browser} and \emph{Tencent Mobile Safeguard}.
Extensive experimental results have demonstrated that FDML can even closely approach centralized learning in terms of testing errors,
without violating data locality constraints,
although centralized learning can use a more sophisticated model, since all features are collected centrally. In the meantime, FDML significantly outperforms models trained only based on the local features of each single app, demonstrating its advantage in harvesting insights from additional cross-domain features.

\section{Related Work}


{\bf Distributed Machine Learning.} Distributed machine learning algorithms and systems have been extensively studied in recent years to scale up machine learning in the presence of big data and big models. 
Existing work focuses either on the theoretical convergence speed of proposed algorithms, or on the practical system aspects to reduce the overall model training time \cite{xing2016strategies}. Bulk synchronous parallel algorithms (BSP) \cite{dekel2012optimal,zinkevich2010parallelized} are among the first distributed machine learning algorithms. Due to the hash constraints on the computation and communication procedures, these schemes share a convergence speed that is similar to traditional synchronous and centralized gradient-like algorithms. Stale synchronous parallel (SSP) algorithms \cite{ho2013more} are a more practical alternative that abandons strict iteration barriers, and allows the workers to be off synchrony up to a certain bounded delay. 
The convergence results have been developed for both gradient descent and SGD \cite{recht2011hogwild,ho2013more, lian2015asynchronous} as well as proximal gradient methods \cite{li2014communication} under different assumptions of the loss functions. In fact, SSP has become central to various types of current distributed Parameter Server architectures \cite{li2014scaling, chilimbi2014project, chen2015mxnet, li2016difacto, abadi2016tensorflow, hsieh2017gaia}. 

Depending on how the computation workload is partitioned \cite{xing2016strategies}, distributed machine learning systems can be categorized into {\it data parallel} and {\it model parallel} systems.
Most of existing distributed machine learning systems \cite{li2014scaling, chilimbi2014project, chen2015mxnet, li2016difacto, abadi2016tensorflow, hsieh2017gaia} fall into the range of {\it data parallel}, where different workers hold different training samples. 

{\bf Model Parallelism.} There are only a couple of studies on \emph{model parallel} systems, i.e., DistBelief \cite{dean2012large} and STRADS \cite{lee2014model}, which aims to train a big model by letting each worker be responsible for updating a subset of model parameters. However, both  DistBelief and STRADS, require collaborating workers to transmit their local model parameters to each other (or to a server), which violates our non-leakage requirement for models and inevitably incurs more transmission overhead. Furthermore, nearly all recent advances on model parallel neural networks (e.g., DistBelief \cite{dean2012large} and AMPNet \cite{ben2018demystifying}) mainly partition the network horizontally according to neural network layers with motivation to scale up computation to big models.
In contrast, we study a completely vertical partition strategy based strictly on features, which is motivated by the cooperation between multiple businesses/organizations that hold different aspects of information about the same samples. Another difference is that we do not require transmitting the model parameters; nor any raw feature data between parties.

On a theoretical perspective of model parallel algorithm analysis, \cite{zhou2016convergence} has proposed and analyzed the convergence of a model parallel yet non-stochastic \emph{proximal gradient} algorithm that requires passing model parameters between workers under the SSP setting. Parallel coordinate descent algorithms have been analyzed recently in \cite{bradley2011parallel,scherrer2012feature}. Yet, these studies focus on randomized coordinate selection in a \emph{synchronous} setting, which is different from our setting where multiple nodes can update disjoint model blocks asynchronously.
Although Stochastic gradient descent (SGD) is the most popular optimization method extensively used for modern distributed data analytics and machine learning, to the best of our knowledge, there is still no convergence result of (asynchronous) SGD in a model parallel setting to date. Our convergence rate of FDML offers the first analysis of asynchronous \emph{model parallel} SGD, which matches the standard convergence rate of the original SSP algorithm \cite{ho2013more} for data parallel SGD.

{\bf Learning Privately.} A variant of distributed SGD with a filter to suppress insignificant updates has recently been applied to collaborative deep learning among multiple parties in a data parallel fashion \cite{shokri2015privacy}. 
Although raw data are not transferred by the distributed SGD in \cite{shokri2015privacy}, a recent study \cite{hitaj2017deep} points out that an algorithm that passes model parameters may be vulnerable to model inversion attacks based on generative adversarial networks (GANs). In contrast, we do not let parties transfer local model parameters to server or any other party.   

Aside from the distributed optimization approach mentioned above, another approach to privacy preserving machine learning is through
encryption, e.g., via homomorphic encryption \cite{gilad2016cryptonets,takabi2016privacy} or secret sharing \cite{mohassel2017secureml, wan2007privacy, bonte2018privacy}. Models are then trained on encrypted data. However, this approach cannot be flexibly generalized to all algorithms and operations, and incurs additional computation and design cost. Relatively earlier, differential privacy has also been applied to collaborative machine learning \cite{pathak2010multiparty, rajkumar2012differentially}, with an inherent tradeoff between privacy and utility of the trained model. To the best of our knowledge, none of the previous work addressed the problem of collaborative learning when the features of each training sample are distributed on multiple participants.

\section{Problem Formulation}
\label{sec:problem}
\begin{figure}[t]
    \centering
    \includegraphics[width=3.2in]{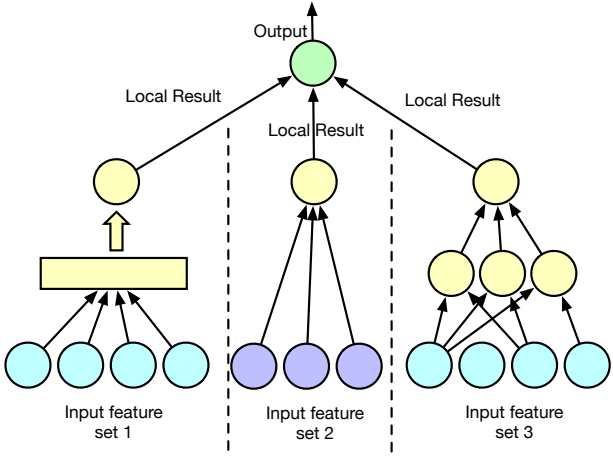}
    \vspace{-4mm}
    \caption{An illustration of the FDML model \eqref{eq:hyper_linear_model}, where each party may adopt an arbitrary local model that is trainable via SGD. The local predictions, which only depend on the local model parameters, are aggregated into a final output using linear and nonlinear transformations \eqref{eq:hyper_linear_model0}.}
    \label{fig:model}
    \vspace{-7mm}
\end{figure}

Consider a system of $m$ different parties, each party holding different aspects about the same training samples. Let $\{(\xi_i^1, \xi_i^2, \ldots, \xi_i^m), y_i\}_{i=1}^n$ represent the set of $n$ training samples, where the vector $\xi_i^j\in\mathbb{R}^{d^j}$ denotes the features of the $i$th sample located on $j$th party, and $y_i$ is the label of sample $i$. 
Let $\xi_i\in\mathbb{R}^d$ be the overall feature vector of sample $i$, which is a concatenation of the vectors $\xi_i^1, \xi_i^2, \ldots, \xi_i^m$, with $d=\sum_j d_j$.
Suppose the parties are not allowed to transfer their respective feature vector to each other out of regulatory and privacy reasons as has been mentioned above. In our problem, the feature vectors on two parties may or may not contain overlapped features. 
The goal of machine learning is to find a model $p(x,\xi)$ with parameters $x$ that given an input $\xi$, can predict its label $y$, by minimizing the loss between the model prediction $p(x,\xi_i)$ and its corresponding label $y_i$ over all training samples $i$.

We propose a Feature Distributed Machine Learning (FDML) algorithm that can train a joint model by utilizing all the distributed features while keeping the raw features at each party unrevealed to other parties.
To achieve this goal, we adopt a specific class of model that has the form
\begin{align}
    p(x, \xi) = \sigma\bigg(\sum_{j=1}^m a_j\alpha^j(x^j, \xi^j)\bigg),\label{eq:hyper_linear_model0}
\end{align}
where $\alpha^j:\mathbb{R}^{D^j}\times\mathbb{R}^{d^j}\rightarrow\mathbb{R}$, $j=1,\ldots,m$, is a sub-model on party $j$ with parameters $x^j\in\mathbb{R}^{D^j}$, which can be a general function that maps the local features $\xi^j$ on each party $j$ to a local prediction. In addition, $\sigma: \mathbb{R}\rightarrow\mathbb{R}$ is a continuously differentiable function to aggregate local intermediate predictions $\alpha^j(x^j, \xi^j)$ weighted by $a_j$. 
Note that $x\in\mathbb{R}^D$, with $D=\sum_j D_j$, is a concatenation of the local model parameters $x^j$ over all parties $j$. 


As illustrated by Fig.~\ref{fig:model}, the model adopted here is essentially a composite model, where each sub-model $\alpha^j$ on party $j$ with parameters $x^j$ could be an arbitrary model, e.g., logistic regression, SVM, deep neural networks, factorization machines, etc. 
Each sub-model $x^j$ on party $j$ is only concerned with the local features $\xi^j$.  
The final prediction is made by merging the local intermediate results through a linear followed by nonlinear transformations, e.g., a softmax function. Note that in \eqref{eq:hyper_linear_model0}, all $a_j$ can be eliminated by scaling some corresponding parameters in $\alpha(x^j, \xi^j)$ by $1/a_j$. Without loss of generality, we simplify the model to the following:
\begin{align}
    p(x, \xi) = \sigma\bigg(\sum_{j=1}^m \alpha ^j(x^j, \xi^j)\bigg),\label{eq:hyper_linear_model}
\end{align}

Apparently, in this model, both the local features $\xi^j$ and the sub-model parameters $x^j$ are stored and processed locally within party $j$, while only the local predictions $\alpha ^j(x^j, \xi^j)$ need be shared to produce the final prediction. Therefore, the raw features as well as all sub-model parameters are kept private. In Sec.~\ref{sec:algorithm}, we propose an asynchronous SGD algorithm that also preserves the non-sharing properties for all the local features as well as all sub-model parameters even during the model training phase, with theoretical convergence guarantees.

In general, the model is trained by solving the following problem: 
\begin{align}
	\text{minimize}_{x} & \frac{1}{n} \sum_{i=1}^n L(x; \xi_i, y_i) + \lambda \sum_{j=1}^m z^{j}(x^j), \label{eq:general_optimization_objective}
\end{align}
where $L\big(p(x, \xi); y\big)$ 
is the loss function, indicating the gap between the predicted value and the true label for each sample. $z(x^j)$ is the regularizer for sub-model $x^j$. 
\section{Asynchronous SGD for FDML}
\label{sec:algorithm}

In this section, we describe our asynchronous and distributed stochastic gradient descent (SGD) algorithm specifically designed to solve the optimization problem \eqref{eq:general_optimization_objective} in FDML, with theoretical convergence guarantees. 

Since we consider a stochastic algorithm, let $i(t)$ be the index of the sample $\xi_{i(t)}$ presented to the training algorithm in iteration $t$. 
To simplify notations, we denote the regularized loss of sample $i(t)$ by
\begin{align}
    F_t(x) := L(x; \xi_{i(t)}, y_{i(t)}) + \lambda \sum_{j=1}^m z^j(x^j), \label{eq:big_F_def}
\end{align} 
Thus, in stochastic optimization, minimizing the loss in \eqref{eq:general_optimization_objective} over the entire training set is equivalently to solving the following problem \cite{ho2013more}:   
\begin{align}
    \text{minimize}_x F(x) := \frac{1}{T}\sum_t F_t(x),
\end{align}
where $T$ is the total number of iterations. Let $\nabla F(x)\in\mathbb{R}^D$ be the gradient of $F$. Let $\nabla^j F(x)\in\mathbb{R}^{D^j}$ be the partial gradient of $F$ with respect to the sub-model parameters $x^j\in\mathbb{R}^{D^j}$, i.e., $\nabla^j F(x) := \frac{\partial F(x)}{\partial x^j}$.
Clearly, $\nabla F(x)$ is the concatenation of all the partial gradients $\nabla ^1F(x), \nabla^2 F(x),\ldots, \nabla^m F(x)$.

\subsection{The Synchronous Algorithm}
In a synchronous setting, we can simply parallelizing a SGD algorithm by updating each parameter block $x^j$ concurrently for all $j=1,\ldots, m$, given a coming sample $i(t)$, i.e.,
$x^j_{t+1} := x^j_t - \eta_t\nabla^j F_t(x^1_t,\ldots,x^m_t)$,
where $\eta_t$ is a predefined learning rate scheme. 
Specifically for model \eqref{eq:hyper_linear_model}, according to \eqref{eq:big_F_def}, we can obtain the partial gradient $\nabla^j F_t(x)$ for $j=1,\ldots,m$ as
\begin{align}
    &\nabla^j F_t(x) = \lambda\frac{\partial z^j(x^j)}{\partial x^j} + \nonumber\\
    & L'\bigg(\sigma\bigg(\sum_{k=1}^m \alpha^k(x^k, \xi^k_{i(t)})\bigg)\bigg)\sigma'\big(\sum_{k=1}^m \alpha^k(x^k, \xi^k_{i(t)})\big)\frac{\partial \alpha^j(x^j, \xi^j_{i(t)})}{\partial x^j} \\
    &:= H\bigg(\sum_{k=1}^m \alpha^k(x^k, \xi^k_{i(t)})\bigg)\frac{\partial \alpha^j(x^j, \xi^j_{i(t)})}{\partial x^j}+ \lambda\frac{\partial z^j(x^j)}{\partial x^j}, \label{eq:partial_gradient_detail}
\end{align}
where we simplify the notation of the first few terms related to $\sum_{k=1}^m \alpha^k(x^k, \xi^k_{i(t)})$ by a function $H(\cdot)$. In practice, $z^j$ could be non-smooth. This setting is usually handled by proximal methods. In this work, we are only focused on the smooth case.

This indicates that for the class of models in \eqref{eq:hyper_linear_model} adopted by FDML, each party $j$ does not even need other parties' models $x^k$, where $k\ne j$, to compute its partial gradient $\nabla^j F_t$. Instead, to compute $\nabla^j F_t$ in \eqref{eq:partial_gradient_detail}, each party $j$ only needs one term, $\sum_{k=1}^m \alpha^k(x^k, \xi^k_{i(t)})$, which is the aggregation of the local prediction results from all parties at iteration $t$, while the remaining terms in \eqref{eq:partial_gradient_detail} is only concerned with party $j$'s local model $x^j$ and local features $\xi^j_{i(t)}$. 
Therefore, this specific property enables a parallel algorithm with minimum sharing among parties, where neither local features nor local model parameters need be passed among parties. 

\subsection{The Asynchronous Algorithm}
The \emph{asynchronous} implementation of this idea in a distributed setting of multiple parties, with theoretical convergence guarantees, is significantly more challenging than it seems.  
As our proposed algorithm is closely related to asynchronous SGD, yet extends it from the data-parallel setting \cite{ho2013more} to a block-wise model parallel setting, we would call our algorithm Asynchronous SGD for FDML.

Note that in an asynchronous setting, each party $j$ will update its own parameters $x^j_t$ asynchronously and two parties may be in different iterations. However, we assume different parties go through the samples $\xi_{i(t)}$ in the same order, although asynchronously, i.e., all the parties share the randomly generated sample index sequence $\{i(t)|t=1,\ldots,T\}$, which can easily be realized by sharing the seed of a pseudo random number generator. 


When each party $j$ has its own iteration $t$, the local model parameters $x_{t}^j$ on party $j$ is updated by 
\begin{align}
    & x_{t+1}^j = x_{t}^j - 
    \eta_t \bigg(H\bigg(\sum_{k=1}^m \alpha^k(x^k_{t-\tau^j_t(k)}, \xi_{i(t)}^k)\bigg)\frac{\partial \alpha^j(x^j_t, \xi^j_{i(t)})}{\partial x^j}+ \lambda\frac{\partial z^j(x^j_t)}{\partial x^j}\bigg), \label{eq:update}
\end{align}
where the requested aggregation of local predictions for sample $\xi_{i(t)}$ may be computed from possibly \emph{stale versions} of model parameters, $x^k_{t-\tau^j_t(k)}$ on other parties $k\ne j$, where $\tau^j_t(k)$ represents how many iterations of a ``lag'' there are from party $k$ to party $j$ at the $t$th iteration of party $j$. We abuse the word ``lag'' here since party $k$ could be faster than party $j$. We overflow the notation for that case by assigning negative value to $\tau^j_t(k)$. We give a convergence speed guarantee of the proposed algorithm under certain assumptions, when the lag $\tau^j_t(k)$ is bounded.

\section{Distributed Implementation} 
\label{sec:implementation}

\subsection{Implementation}
We present a distributed implementation of the proposed asynchronous SGD algorithm for FDML. Our implementation is inspired by the Parameter Server architecture \cite{li2014scaling, li2014communication, chilimbi2014project}.
In a typical Parameter Server system, the workers compute gradients while the server updates the model parameters with the gradients computed by workers. Yet, in our implementation, as described in Algorithm~\ref{alg:FDML}, the only job of the server is to maintain and update a matrix $A_{i,j}$, $i=1,\ldots,n$, $j=1,\ldots,m$, which is introduced to hold the latest $m$ local predictions for each sample $i$. We call $[A_{i,j}]_{n\times m}$ the \emph{local prediction matrix}. On the other hand, unlike servers, the workers in our system each represent a participating party. They do not only compute gradients, but also need to update their respective local model parameters with SGD. 

Furthermore, since each worker performs local updates individually, each worker can even further employ a parameter server cluster or a shared-memory system, e.g., a CPU/GPU workstation, to scale up and parallelize the computation workload related to any local model it adopts, e.g., a DNN or FM. A similar hierarchical cluster is considered in Gaia \cite{hsieh2017gaia}, though for data-parallel machine learning among multiple data centers.

\begin{algorithm}[t]
\caption{A Distributed Implementation of FDML}
\hspace*{\algorithmicindent} \textbf{Require}: each {\tt worker} $j$ holds the local feature set $\{\xi_i^j,y_i\}_{i=1}^n$, $j=1,\ldots,m$; a sample presentation schedule $i(t)$, $t=1,\ldots,T$, is pre-generated randomly and shared among workers. \\
\hspace*{\algorithmicindent} \textbf{Output}: model parameters $x_T = (x_T^1,\ldots,x_T^m)$.
\vspace{-5mm}
\begin{algorithmic}[]
\STATE {\tt \textbf{Server}}:
\bindent
\STATE Initialize the local prediction matrix $[A_{i,j}]_{n\times m}$.
\WHILE {True}
	\IF {{\it Pull request} (worker: $j$, iteration: $t$) received }
		\IF {$t$ is not $\tau$ iterations ahead of the slowest worker}
			\STATE Send $\sum_{k=1}^m A_{i(t),k}$ to {\tt{Worker}} $j$
		\ELSE
			\STATE Reject the \emph{Pull request}
		\ENDIF
	\ENDIF
	\IF {{\it Push request} (worker: $j$, iteration: $t$, value: $c$) received}
		\STATE $A_{i(t),j}:= c$.
	\ENDIF
\ENDWHILE
\eindent 
\STATE {\tt \textbf{Worker}} $j$ ($j=1,\ldots,m$) asynchronously performs:
\bindent
	\FOR {$t=1,\ldots, T$}
		\STATE Push $c := \alpha^j(x_{t}^j, \xi_{i(t)}^j)$ to {\tt {Server}}
		\WHILE {Pull not successful}
			\STATE Pull $\sum_{k=1}^m A_{i(t),k}$ from {\tt {Server}}
		\ENDWHILE
		\STATE $\nabla^j F_t := \bigg(H(\sum_{k=1}^m A_{i(t),k})\cdot\frac{\partial \alpha^j(x^j_t, \xi^j_{i(t)})}{\partial x^j}+ \lambda \frac{\partial z^j(x^j_t)}{\partial x^j}\bigg)$
		\STATE Update the local weights as {\small
		\begin{align}
			& x_{t+1}^j := x_t^j - \eta_t \nabla^j F_t\label{eq:x_t_alg}.
		\end{align}
		}
	\ENDFOR
\eindent
\end{algorithmic}
\label{alg:FDML}
\end{algorithm}

First, we describe how the input data should be prepared for the FDML system. Before the training task, for consistency and efficiency, a {\it sample coordinator} will first randomly shuffle the sample indices and generate the sample presentation schedule $i(t)$, which dictates the order in which samples should be presented to the training algorithm.
However, since features of a same sample are located on multiple parties, we need to find all the local features $\xi_i^1, \xi_i^2, \ldots, \xi_i^m$ as well as the label $y_i$ associated with sample $i$. This can be done by using some common identifiers that are present in all local features of a sample, like user IDs, phone numbers, data of birth plus name, item IDs, etc. Finally, the labels $y_i$ will be sent to all workers (parties) so that they can compute error gradients locally. Therefore, before the algorithm starts, each worker $j$ holds a local dataset $\{\xi_i^j,y_i\}_{i=1}^n$, for all $j=1,\ldots,m$.

Let us explain Algorithm~\ref{alg:FDML} from a worker's perspective. 

To solve for $x$ collaboratively, each worker $j$ goes through the iterations $t=1,\ldots,T$ individually and asynchronously in parallel, according to the (same) predefined sample presentation schedule $i(t)$ and updates its local model $x^j$ according to \eqref{eq:x_t_alg}.
In a particular iteration $t$, when worker $j$ updates $x_t^j$ with the current local features $\xi_{i(t)}^j$, it first sends its updated local prediction about sample $i(t)$ to the server in order to update $A_{i(t),j}$, i.e., $A_{i(t),j} := \alpha^j(x_{t}^j, \xi_{i(t)}^j)$. And this update is done through the value $c$ uploaded to the server in a \emph{Push request} from worker $j$ with iteration index $t$ and value $c$. After this update, it pulls the latest $\sum_{k=1}^m A_{i(t),k}$ from the server based on the latest versions of local predictions, $A_{i(t),k}$, maintained on the server for all the workers $k=1,\ldots,m$. Then $x_t^j$ is updated into $x_{t+1}^j$ locally by \eqref{eq:x_t_alg}.



Since the workers perform local model updates asynchronously, at a certain point, different workers might be in different iterations, and a faster worker may be using the stale local predictions from other workers. 
We adopt a \emph{stale synchronous} protocol to strike a balance between the evaluation time for each iteration and the total number of iterations to converge---a fully synchronous algorithm takes the least number of iterations to converge yet incurs large waiting time per iteration due to straggler workers, while on the other hand, an asynchronous algorithm reduced the per iteration evaluation time, at the possible cost of more iterations to converge. 
In order to reduce the overall training time, we require that the iteration of the fastest party should not exceed the iteration of the slowest party by $\tau$, i.e., the server will reject a pull request if the $t$ from the \emph{Pull request}(worker: $j$, iteration: $t$) is $\tau$ iterations ahead of the slowest worker in the system. A similar bounded delay condition is enforced in most Parameter-Server-like systems \cite{li2014scaling, chilimbi2014project, chen2015mxnet, li2016difacto, abadi2016tensorflow, hsieh2017gaia} to ensure convergence and avoid chaotic behavior of a completely asynchronous system.


In real applications, the SGD algorithm can easily be replaced with the mini-batched SGD, by replacing the sample presentation schedule $i(t)$ with a set $I(t)$ representing the indices of a mini-batch of samples to be used iteration $t$, and replacing the partial gradient in \eqref{eq:update} with the sum of partial gradients over the mini-batch $I(t)$.


\subsection{Privacy}

In FDML, one of the primary concerns is to preserve the privacy of the local feature data. Due to the specific model structure and the well designed algorithm, no model weights or features are uploaded from any parties. The only shared information is the intermediate local prediction results for each training or testing sample, which is some comprehensive function over both the local features and model weights. Therefore, there is little chance to leak the original features to honest servers or other parties. 

To further protect the feature data at each party from malicious servers and parties, we apply differential privacy based methods by perturbing the local predictions to be uploaded \cite{dwork2008differential, dwork2014algorithmic, shokri2015privacy}. 
In particular, we add some noise to the local prediction result $\alpha^j(x_{t}^j, \xi_{i(t)}^j)$ at party $j$ to protect the privacy of all the input features at party $j$.

\section{Convergence Analysis}

Inspired by a series of studies \cite{langford2009slow, ho2013more, hsieh2017gaia} on the convergence behavior of convex objective functions, we analyze the convergence property of the proposed asynchronous algorithm by evaluating a \emph{regret} function, which is the difference between the aggregated training loss and the loss of the optimal solution, i.e., the regret $R$ is defined as
\begin{align}
    R = \frac{1}{T}\sum_{t} F_t(x_t) - F(x_*),\label{eq:def_of_R}
\end{align}
where $x_*$ is the optimal solution for $F(x)$, such that $x_* = \text{argmin}_x F(x)$.
During training, the same set of data will be looped through for several \emph{epochs}. This is as if a very large dataset is gone through till $T$th iteration.
We will prove convergence by showing that $R$ will decrease to $0$ with regard to $T$. Before presenting the main result, we introduce several notations and assumptions. 
We use $D_t$ to denote the distance measure from  $x_t$ to $x_*$, i.e., $D_t := \frac{1}{2}\left\lVert x_t - x_* \right\rVert^2_2$. We make the following common assumptions on the loss function, which are used in many related studies as well.

 \begin{assumption}\label{theorem:assumption}
    \begin{enumerate}
        \item The function $F_t$ is differentiable and the partial gradient $\nabla^j f$ are {\it Lipschitz continuous} with $L_j$, namely, 
            \begin{align}
                \|\nabla^j F_t(x_1) - \nabla^j F_t(x_2) \| \le L_j\| x_1 - x_2 \|,
            \end{align}
        for $\forall x_1, x_2 \in\mathbb{R}^D$.
        We denote $L_{\text{max}}$ as the maximum among the $L_j$ for $\forall j$.
        \item Convexity of the loss function $F_t(x)$.
        \item Bounded solution space. There exists a $D>0$, s.t., $D_t \le \frac{1}{2} D^2$
            for $\forall t$.
    \end{enumerate}
\end{assumption}
As a consequence of the assumptions, the gradients are bounded, i.e., $\exists G>0$, s.t., $\left\lVert\nabla F(x)\right\rVert^2_2 \le G^2$。
for $\forall x \in\mathbb{R}^D$
With these assumptions, we come to our main result on the convergence rate of the proposed SGD algorithm. 
\begin{proposition}\label{prop:lr_ssp_convergence}
    Under circumstances of the assumptions in Assumption~\ref{theorem:assumption}, with a learning rate of $\eta_t = \frac{\eta}{\sqrt{t}}$, and a bounded staleness of $\tau$, the regret $R$ given by the updates \eqref{eq:update} for the FDML problem is 
    $R=O(\frac{1}{\sqrt{T}})$.  
\end{proposition}

{\bf Proof.} Please refer to Appendix for the proof. 

\section{Experiments}

\begin{figure*}[t]
    \centering
    \vspace{-5mm}
    \subfigure[Training objective vs. epoch]{
    \includegraphics[width=1.7in]{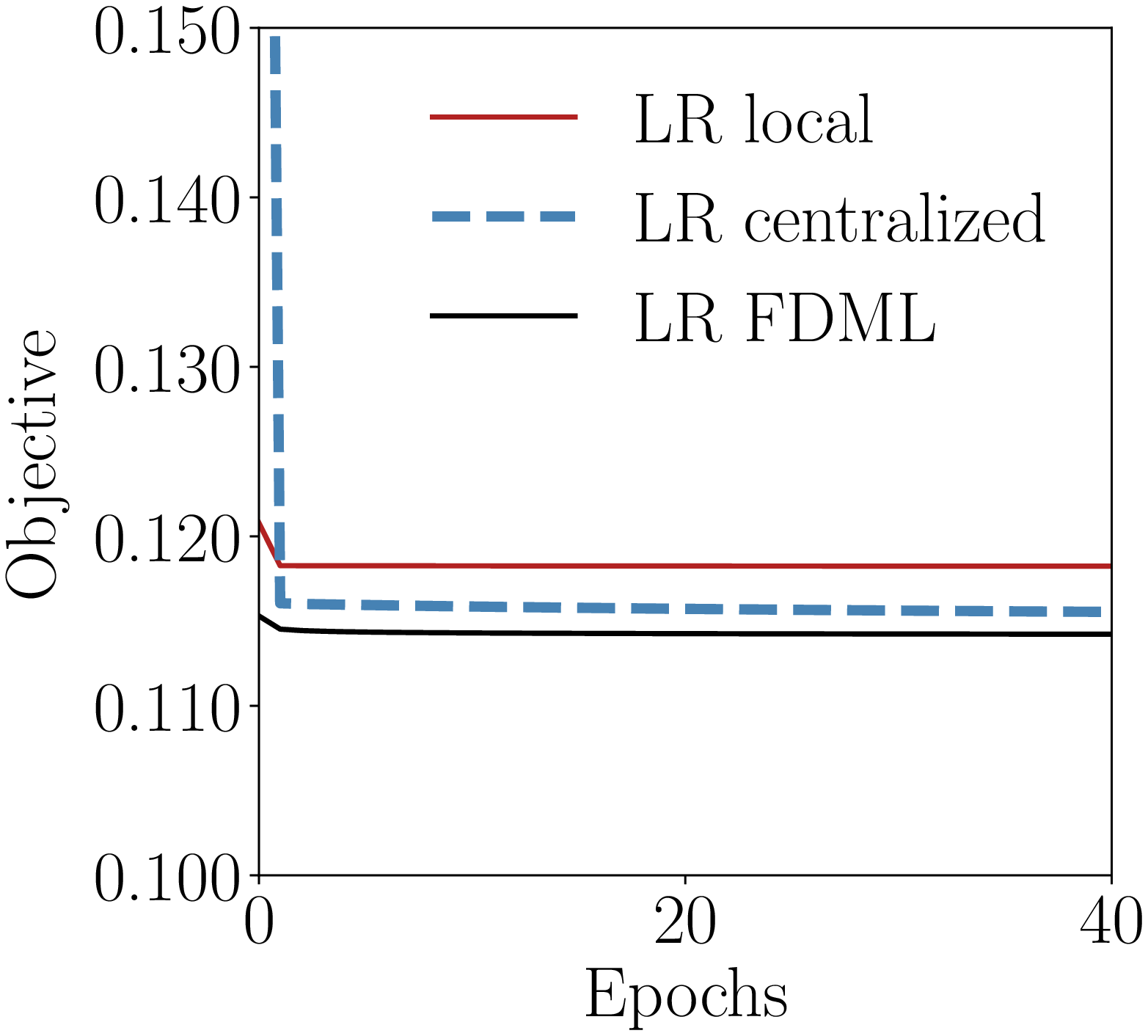}
    \label{fig:LR_perform:train_obj_vs_epoch}
    }
    \hspace{-3.5mm}
    \subfigure[Tesiting log loss vs. epoch]{
    \includegraphics[width=1.7in]{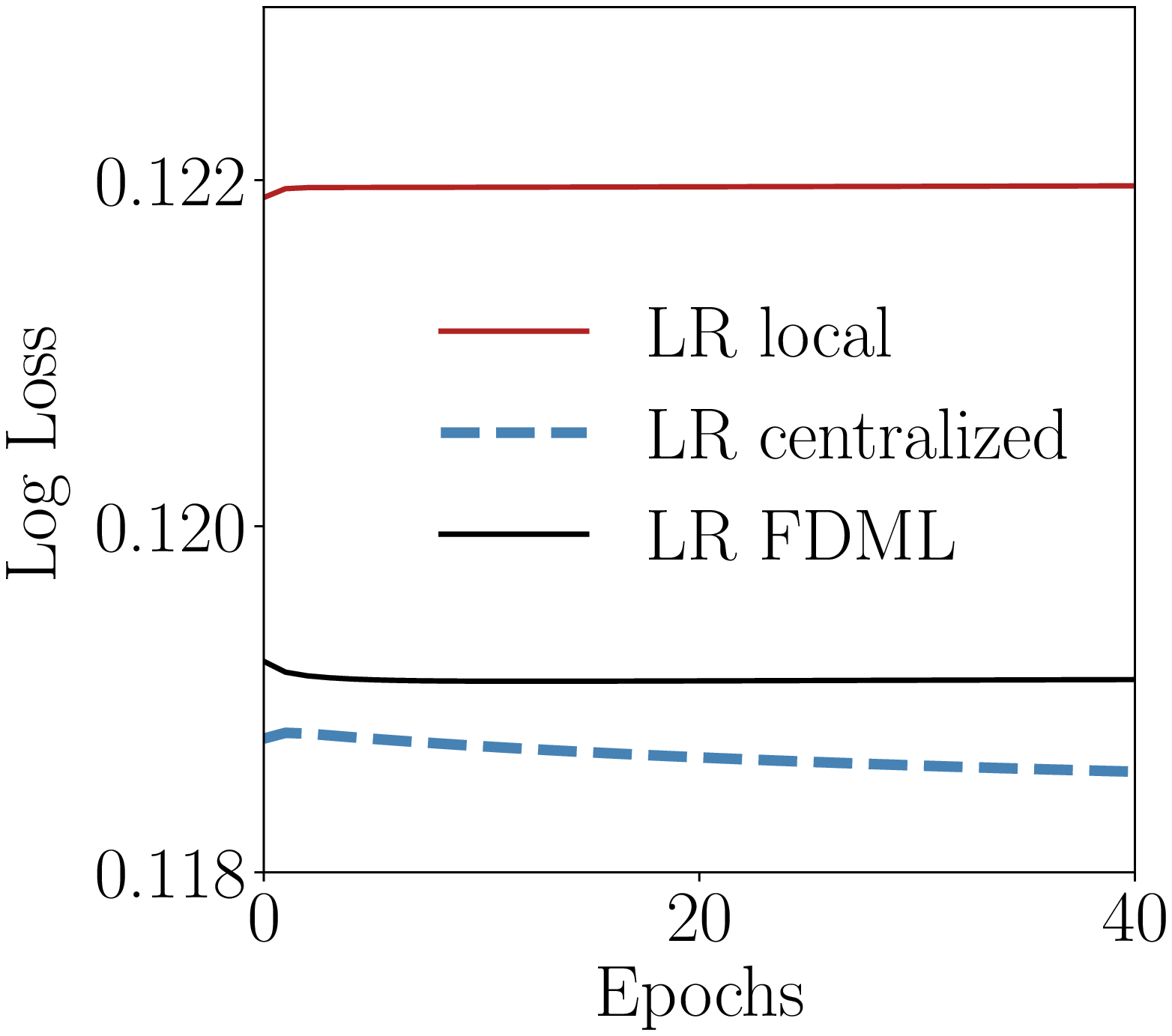}
    \label{fig:LR_perform:test_loss_vs_batch}
    }
    \hspace{-3.5mm}
    \subfigure[Tesiting AUC vs. epoch]{
    \includegraphics[width=1.7in]{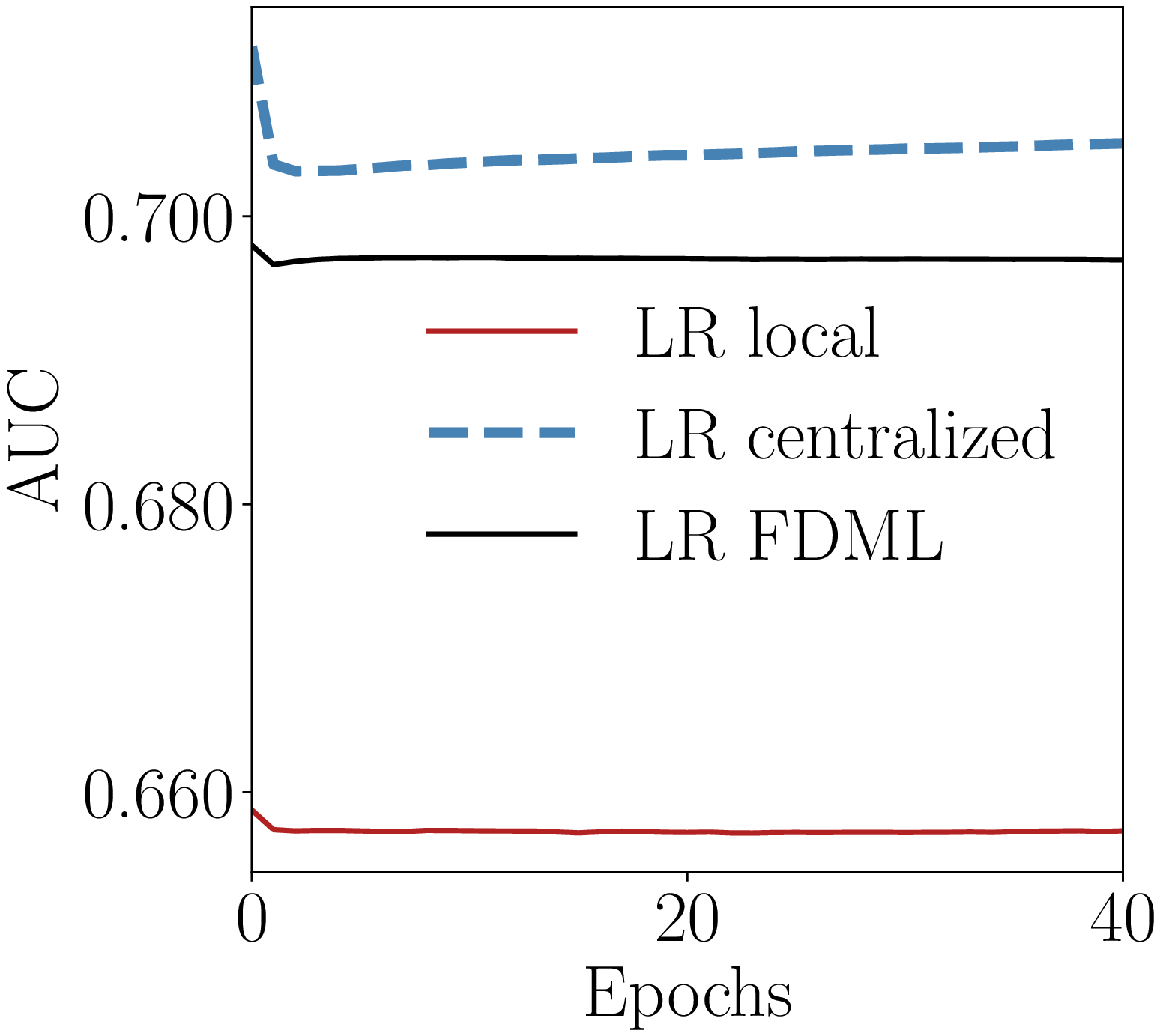}
    \label{fig:LR_perform:test_auc_vs_batch}
    }
    \hspace{-3.5mm}
    \subfigure[Training objective vs. time]{
    \includegraphics[width=1.7in]{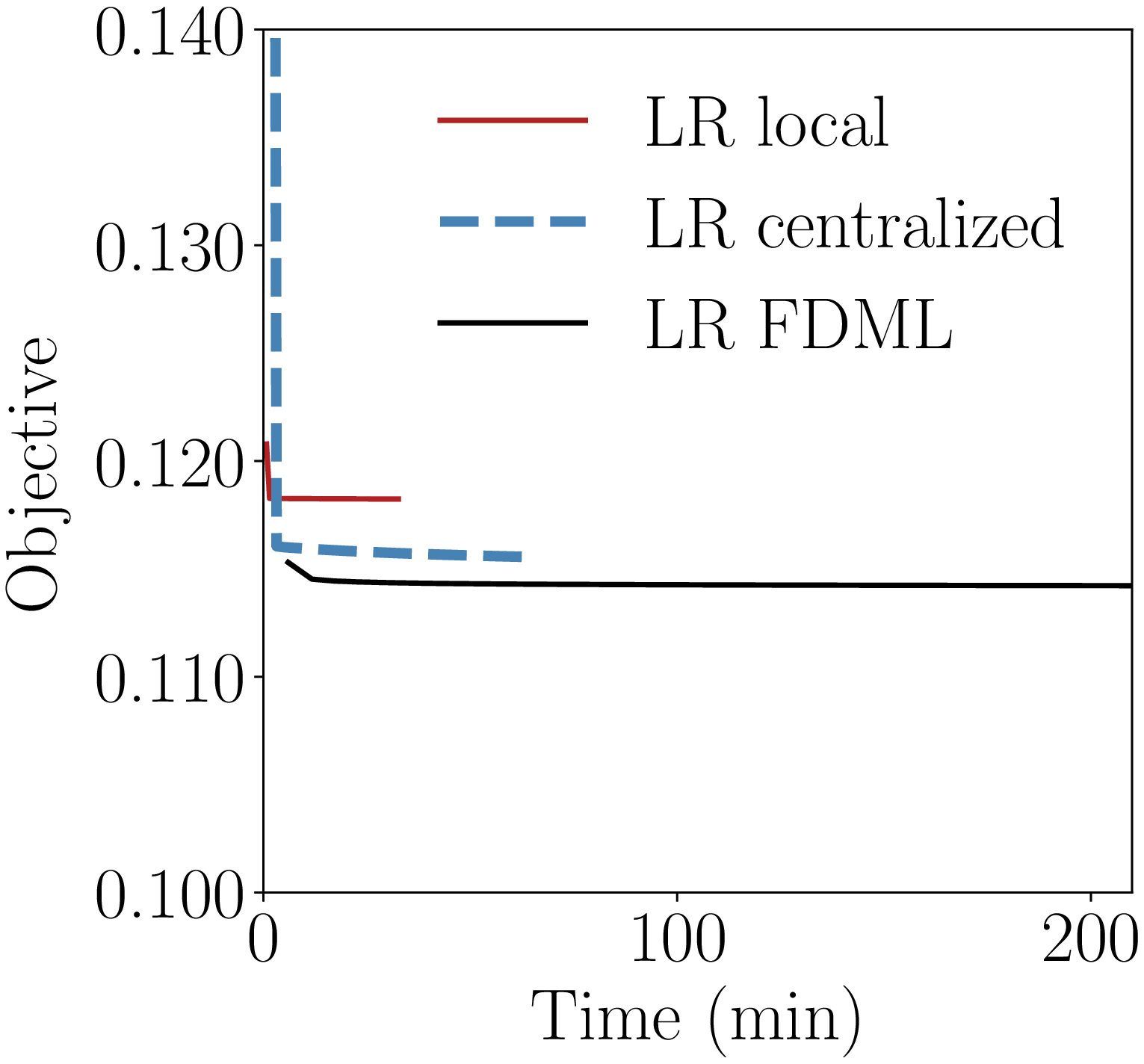}
    \label{fig:LR_perform:train_obj_vs_time}
    }
    \vspace{-3.5mm}
    \caption{A comparison between the three model training schemes for the LR model. All curves are plotted for epochs 1--40, including the time curve in (d).}
    \label{fig:LR_perform}
    \vspace{-4mm}
\end{figure*}

\begin{figure*}[t]
    \centering
    \subfigure[Training objective vs. epoch]{
    \includegraphics[width=1.7in]{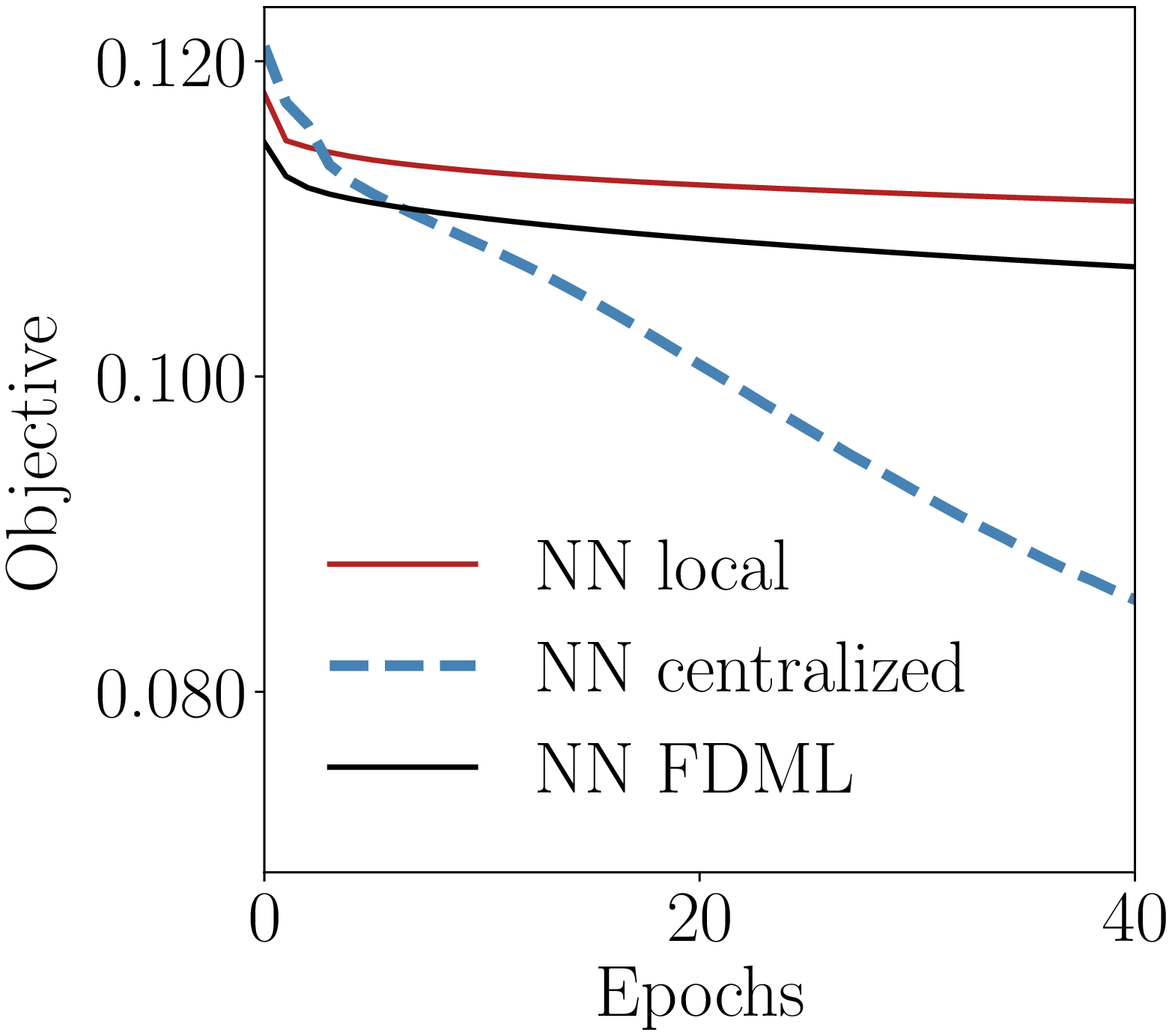}
    }
    \hspace{-3.5mm}
    \subfigure[Tesiting log loss vs. epoch]{
    \includegraphics[width=1.7in]{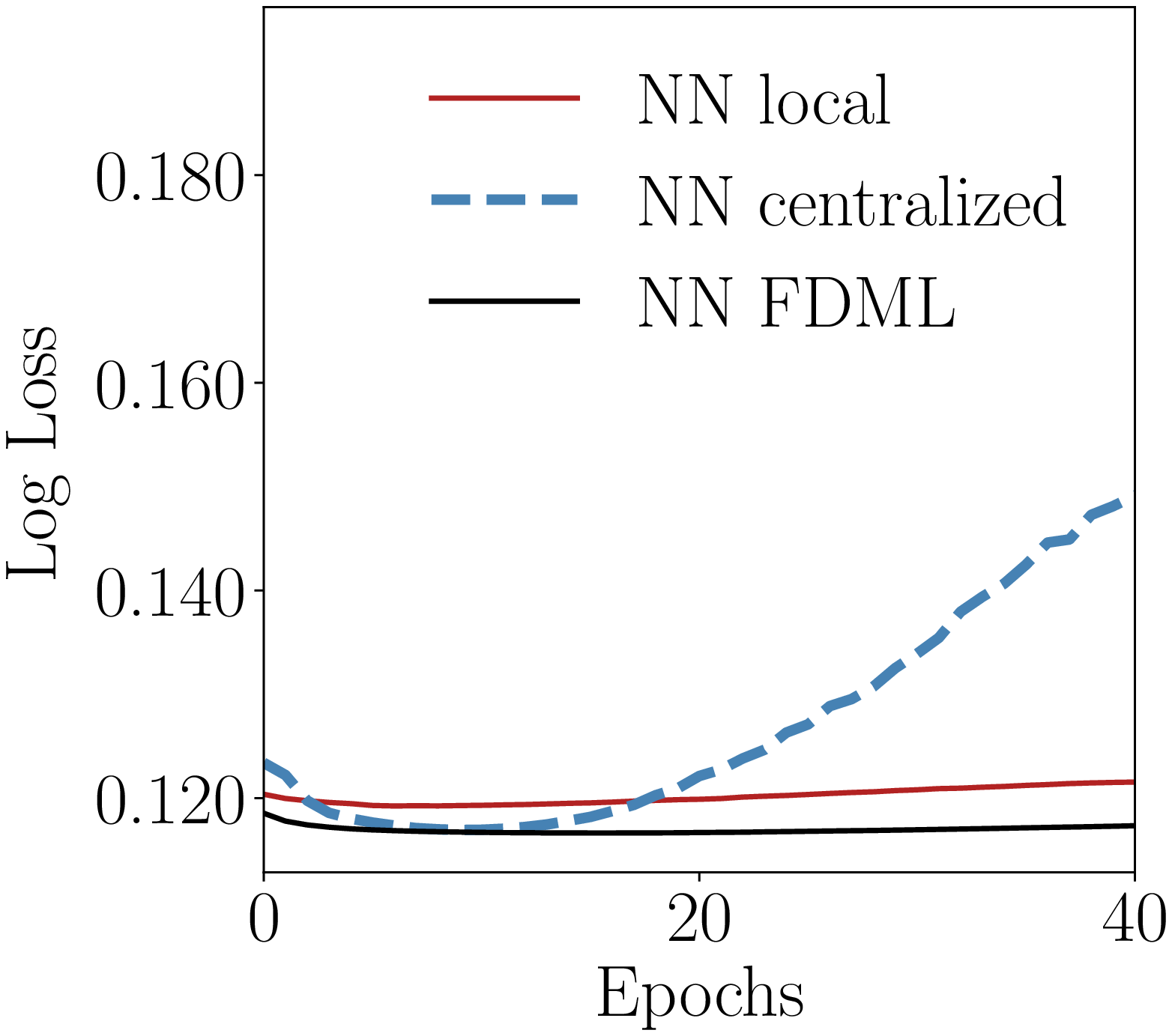}
    }
    \hspace{-3.5mm}
    \subfigure[Tesiting AUC vs. epoch]{
    \includegraphics[width=1.7in]{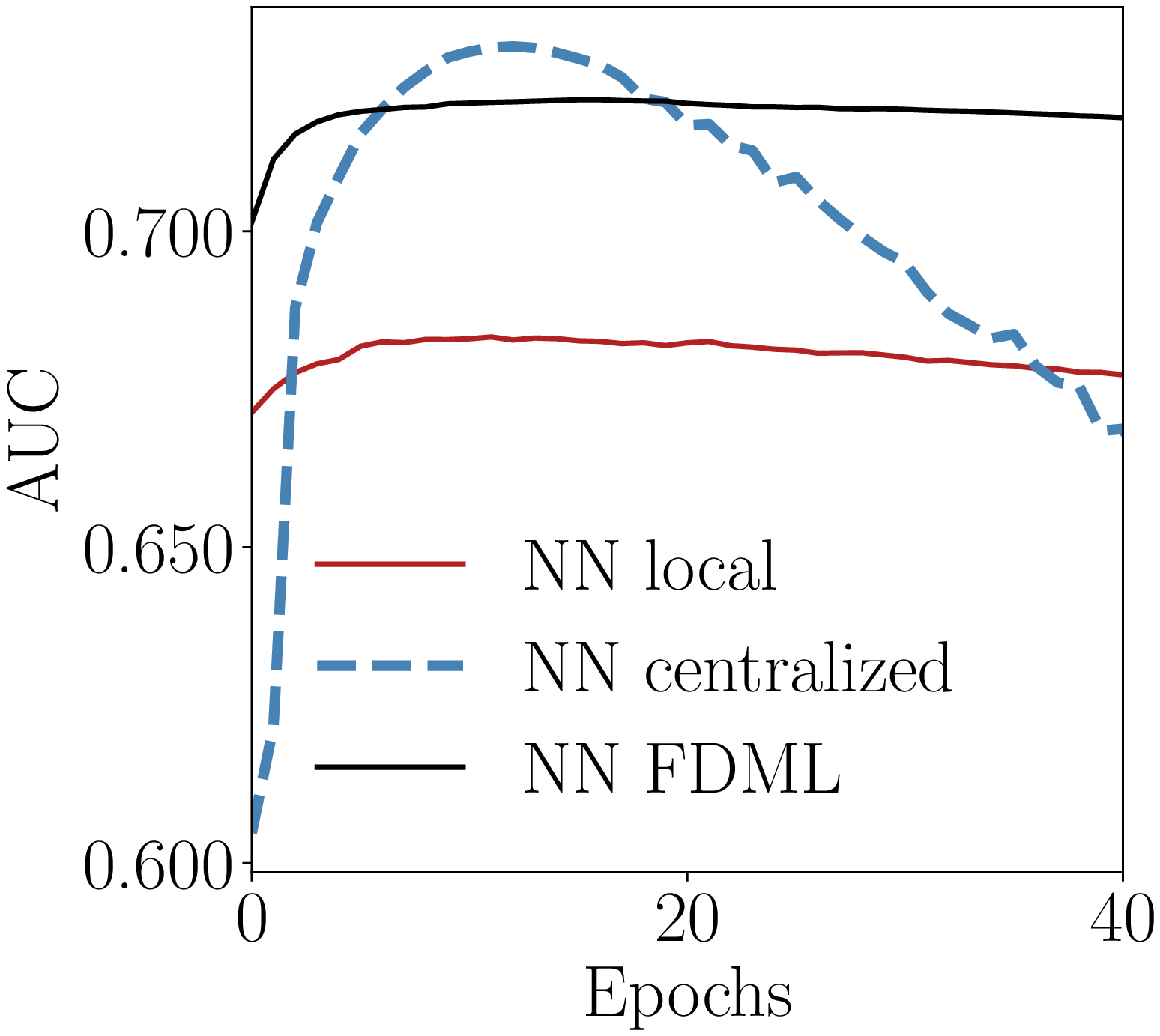}
    }
    \hspace{-3.5mm}
    \subfigure[Training objective vs. time]{
    \includegraphics[width=1.7in]{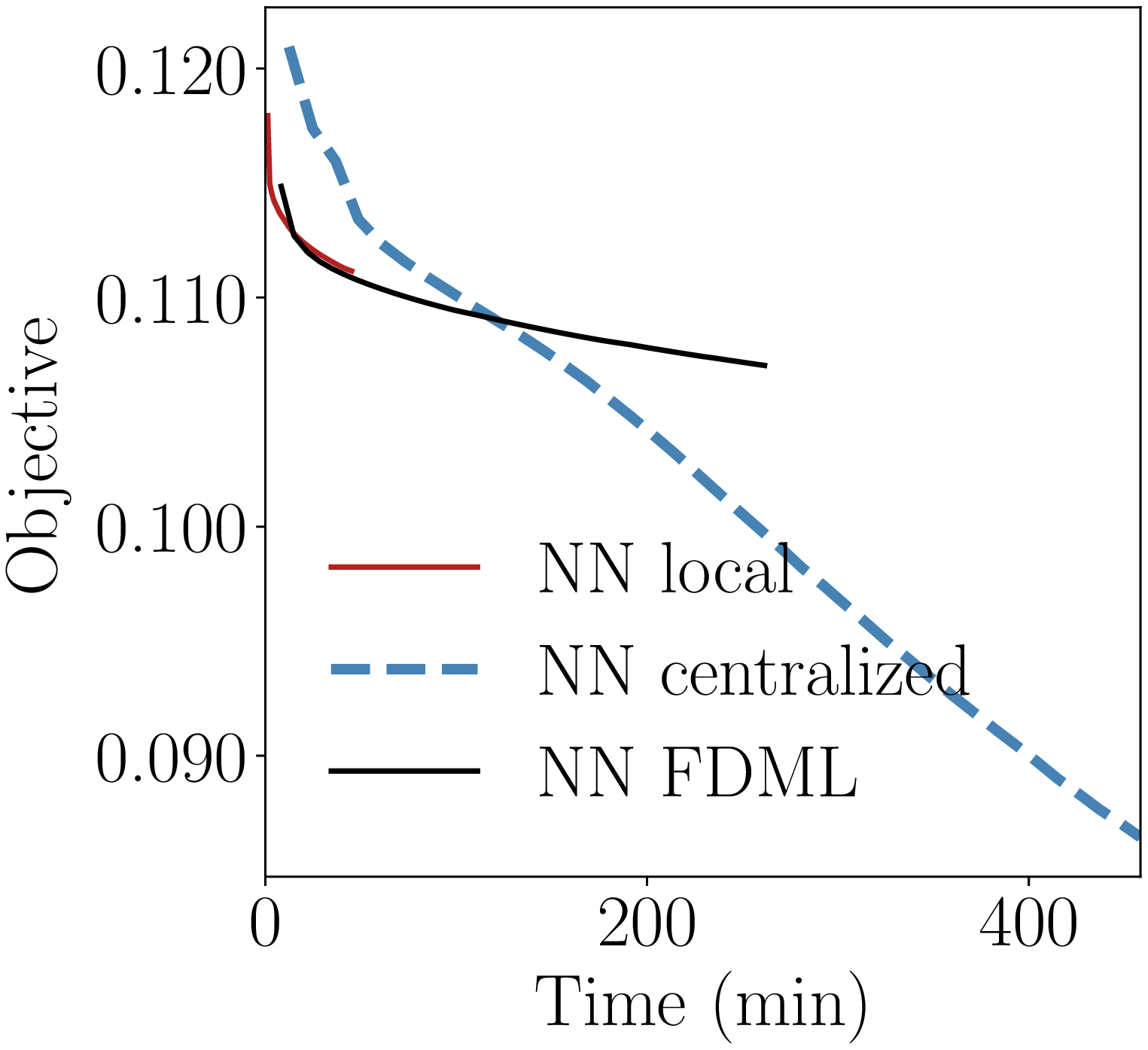}
    \label{fig:dnn_perform:train_obj_vs_time}
    }
    \vspace{-3.5mm}
    \caption{A comparison between the three model training schemes for the NN model. All curves are plotted for epochs 1--40, including the time curve in (d).}
    \label{fig:dnn_perform}
    \vspace{-4mm}
\end{figure*}

\begin{figure}[t]
    \centering
    \subfigure[Test AUC vs. added noise for LR]{
    \includegraphics[width=1.65in]{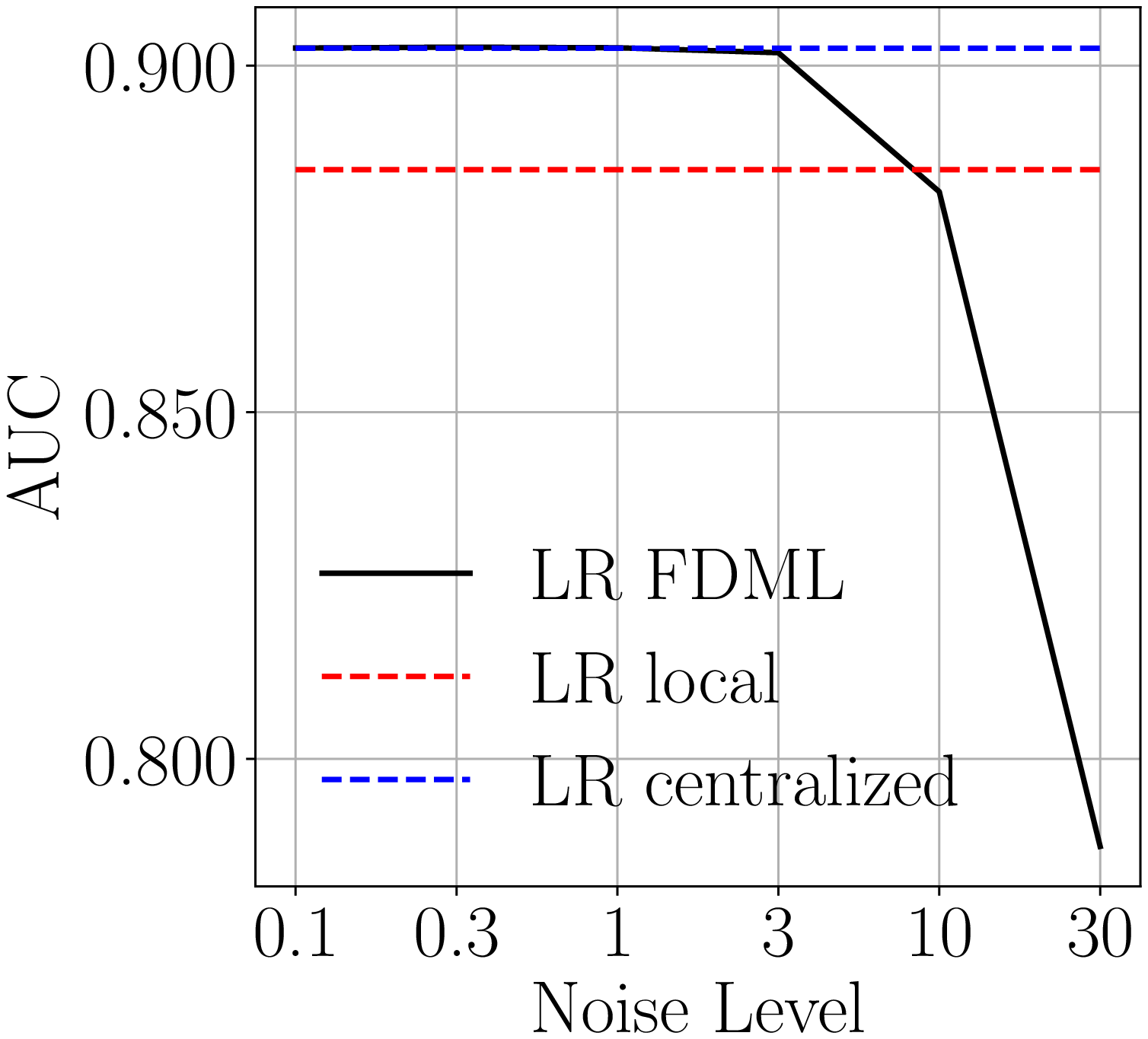}
    }
    \hspace{-4mm}
    \subfigure[Test AUC vs. added noise for NN]{
    \includegraphics[width=1.65in]{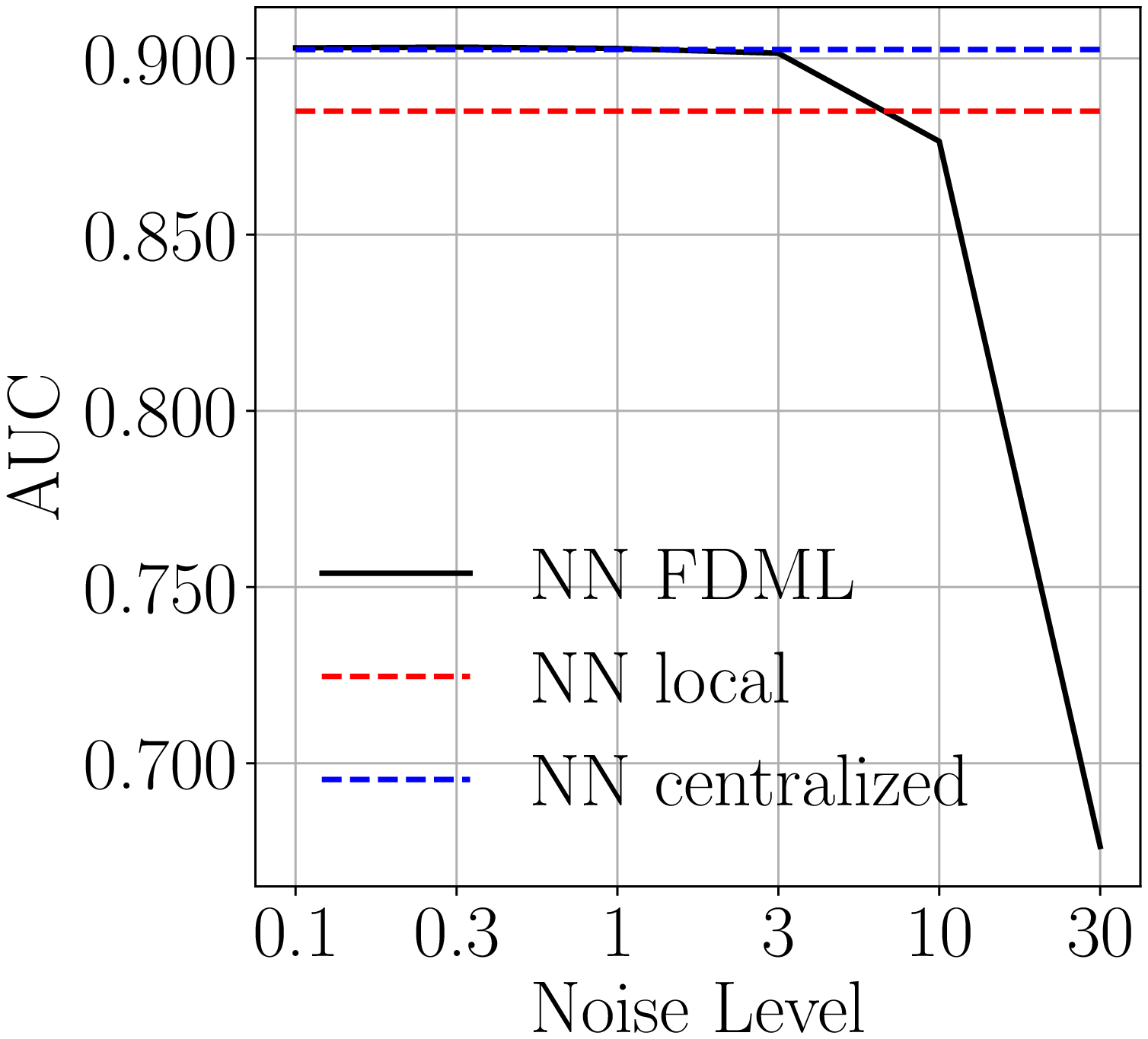}
    }
    \vspace{-4mm}
    \caption{Testing AUC under different levels of added noise during training for {\tt a9a} data set}
    \label{fig:a9a_dp}
    \vspace{-4mm}
\end{figure}

We are testing the application of the proposed FDML system in an app recommendation task at Tencent MyApp, which is a major Android market with an extremely large body of users in China.
In this task, {\it user features}, including the past download activities in MyApp, are recorded. In the meantime, the task can also benefit from cross-domain features about the same users logged in two other apps (run by different departments of the same company), including QQ Browser that tracks user interests in different types of content based on their content viewing history, as well as Tencent Mobile Safeguard, which records the app invoking and usage history of users.

\begin{table}[t]
\caption{The performance of different algorithms for Tencent MyApp data.}
\vspace{-3mm}
\begin{tabular}{lllll}
\hline
Algorithm      & Train loss      & Test loss   & Test AUC        & Time(s) \\ 
\hline
LR local       & 0.1183          & 0.1220          & 0.6573          & 546  \\
LR centralized & 0.1159          & \textbf{0.1187} & \textbf{0.7037} & 1063 \\
LR FDML        & \textbf{0.1143} & 0.1191          & 0.6971          & 3530 \\ \hline
NN local       & 0.1130          & 0.1193          & 0.6830          & 784  \\
NN centralized & \textbf{0.1083} & 0.1170          & \textbf{0.7284} & 8051 \\
NN FDML        & 0.1101          & \textbf{0.1167} & 0.7203          & 4369 \\ \hline
\end{tabular}
\label{tab:main_results}
\vspace{-3mm}
\end{table}

\begin{table}[t]
\caption{The performance of different algorithms for a9a data.}
\vspace{-3mm}
\begin{tabular}{lllll}
\hline
Algorithm      & Train loss      & Test loss   & Test AUC        & Time(s) \\ 
\hline
LR local       & 0.3625 & 0.3509 & 0.8850 & 41 \\
LR centralized & 0.3359 & 0.3247 & 0.9025 & 45 \\
LR FDML        & \textbf{0.3352} & \textbf{0.3246} & \textbf{0.9026} & 99 \\ \hline
NN local       & \textbf{0.3652} & 0.3484 & 0.8864 & 53 \\
NN centralized & 0.4008 & \textbf{0.3235} & \textbf{0.9042} & 57 \\
NN FDML        & 0.4170 & 0.3272 & 0.9035 & 110 \\ \hline
\end{tabular}
\label{tab:a9a_results}
\vspace{-3mm}
\end{table}

The goal here is to leverage the additional user features available from the other domains to improve the app recommendation in MyApp, yet without having to download the raw user features from other apps to avoid regulatory issues, as customer data in different departments are protected under different security levels and even under different customer agreements. Some sensitive features under strong protection are prohibited to be moved to other parties, including other departments.

The dataset we use contains $5,000,000$ labeled samples indicating whether a user will download an app or not. Each sample is a user-app pair, which contains around $8,700$ (sparse) features in total, among which around $7,000$ features come from Tencent MyApp itself, while the remaining $1,700$ features are from the other two apps. We randomly shuffle the data and split it into a 4.5 million training set and a 0.5 million testing set.

We also evaluate FDML on another public data set {\tt a9a} \cite{Dua:2017}, a classical census dataset, where the prediction task is to determine whether a person makes over \$50K a year.
There are $48,842$ samples, each with $124$ features. $32,661$ samples are training data and $16,281$ samples are testing data. We split the $124$ features into two sets of $67$ and $57$. 
We run both a logistic regression (LR) and a two layered fully connected neural network (NN) under three different training schemes for both data sets:
\begin{itemize}
    \item \emph{Local}: only use the $7,000$ local features from MyApp or the $67$ features of {\tt a9a} to train a model.
    \item \emph{Centralized}: collect all the $8,700$ features from all three apps to a central server or using all the $124$ features in {\tt a9a} and train the model using the standard mini-batched SGD. 
    \item \emph{FDML}: use FDML system to train a joint model for app recommendation based on all $8,700$ features distributed in three apps or train the {\tt a9a} classification model on all $124$ features from two different parties, without centrally collecting data. 
\end{itemize}

For FDML, there is a single server with several workers, each of which equipped with an Intel Xeon CPU E5-2670 v3 @ 2.30GHz. Each worker handles the features from one party. The system will be asynchronous as the lengths of features handled by each worker are different.
The FDML NN only considers a fully connected NN within each party while merging the three local predictions in a composite model, whereas the Centralized NN uses a fully connected neural network over all the $8,700$\/$124$ features, thus leading to a more complex model (with interactions between the local features of different departments) than FDML NN. 

For all training schemes, a mini-batched SGD is used with a batch size of 100. For each epoch, we keep track of the optimization objective value for training data, the log loss and the AUC for testing data as long as the elapsed time of the epoch. Fig.~\ref{fig:LR_perform} and Fig.~\ref{fig:dnn_perform} present the major statistics of the models during the training procedure for LR and NN for Tencent MyApp dataset, respectively. Table~\ref{tab:main_results} presents the detailed statistics at the epoch when all the algorithms yield a stable and good performance on the testing data. Table~\ref{tab:a9a_results} presents the performance for {\tt a9a} dataset. The results show that FDML outperforms the corresponding Local scheme with only local features, and even approaches the performance of the Centralized scheme, while keeping the feature sets local to their respective workers.

For LR, as shown by Fig.~\ref{fig:LR_perform}, Table~\ref{tab:main_results} and Table~\ref{tab:a9a_results}, we can see that Centralized LR and FDML LR both achieve a smaller training objective value as well as significantly better performance on the testing set than Local LR. As we have expected, additional features recorded by other related services could indeed help improve the app recommendation performance. Furthermore, Centralizd LR and FDML LR have very close performance, since these two methods use the essentially the same model for LR, though with different training algorithms.

For NN shown in Fig.~\ref{fig:dnn_perform}, Table~\ref{tab:main_results} and Table~\ref{tab:a9a_results}, by leveraging additional features, both FDML NN and Centralized NN substantially outperform Local NN. Meanwhile, Centralized NN is slightly better than FDML NN, since Centralized NN has essentially adopted a more complex model, enabling feature interaction between different parties directly through fully connected neural networks.

Fig.~\ref{fig:LR_perform:train_obj_vs_time} and Fig.~\ref{fig:dnn_perform:train_obj_vs_time} compare the training time and speed among the three learning schemes for Tencent MyApp dataset. Without surprise, for both the LR and NN model, the Local scheme is the fastest since it uses the smallest amount of features and has no communication or synchronization overhead. For LR in Fig.~\ref{fig:LR_perform:train_obj_vs_time}, FDML LR is slower than Centralized LR since the computation load is relatively smaller in this LR model and thus the communication overhead dominates. On the contrary, for NN, as shown in Fig.~\ref{fig:dnn_perform:train_obj_vs_time}, the Centralized NN is slower than FDML NN. This is because Centralized NN has much more inner connections and hence much more model parameters to train. Another reason is that FDML distributes the heavy computation load in this NN scenario to three different workers, which in fact speeds up training. Interestingly, for the smaller dataset {\tt a9a}, in Table~\ref{tab:a9a_results}, the NN FDML is slower than the centralized one since in this case, the model is small and the communication overhead dominate the processing time in FDML. 

Fig.~\ref{fig:a9a_dp} shows the performance when different levels of noise is added according to the differential privacy mechanism 
during the training procedure for {\tt a9a} dataset. 
Conforming to the intuition, we can see that a higher level of noise will bring worse results. However, for a noise level no more than 3, we can still expect a performance improvement over learning only based on local data, while achieving stronger privacy guarantee due to the perturbations introduced to the shared local prediction results.

\vspace{-3mm}
\section{Conclusions}
\label{sec:conclude}

We study a feature distributed machine learning (FDML) problem motivated by real-world recommender applications at Tencent MyApp, where the features about the same training sample can be found at three different apps. However, the features of one app should be kept confidential to other parties due to regulatory constraints. This motivation is in contrast to most existing literature on collaborative and distributed machine learning which assumes the data samples (but not the features) are distributed and works in a data-parallel fashion. We propose an asynchronous SGD algorithm to solve the new FDML scenario, with a convergence rate of $O(1/\sqrt{T})$, $T$ being the total number of iterations, matching the existing convergence rate known for data-parallel SGD in a stale synchronous parallel setting \cite{ho2013more}. 

We have developed a distributed implementation of the FDML system in a parameter server architecture and performed extensive evaluation based on both a public data set and a large dataset of $5,000,000$ records and $8,700$ decentralized features from \emph{Tencent MyApp}, \emph{Tencent QQ Browser} and \emph{Tencent Mobile Safeguard} for a realistic app recommendation task.Results have shown that FDML can closely approximate centralized training (the latter collecting all data centrally and using a more complex model allowing more interactions among cross-domain features) in terms of the testing AUC and log loss, while significantly outperforming the models trained only based on the local features of \emph{MyApp}. Currently, we are deploying the FDML system at \emph{Tencent MyApp} and improving the robustness of the system by adding momentum based techniques. We are also developing schemes that can support more sophisicated models, taking more interactions between cross-party features into account. 

\bibliographystyle{ACM-Reference-Format}
\bibliography{main}
\section{Appendix}

{\bf Proof of Proposition~\ref{prop:lr_ssp_convergence}. }
By the proposed algorithm and from \eqref{eq:update}, we have$x^j_{t+1} = x^j_t - \eta_t \nabla^j F(\tilde{x}_t(j))$,
where $\tilde{x}_t(j)$ is the concatenated model parameters with staleness in which $\tilde{x}^i_t(j) = x^i_{t - \tau^j(i)}$.
Note that we always have $\tau^j(i)\le\tau, \forall i,j$. 
To help proving the proposition, we first prove a lemma. 
\begin{lemma}\label{lamma:main}
\begin{align}
    & <x_t-x_*, \nabla F_t(x_t)> = \frac{1}{2}\eta_t \sum_{j=1}^{m} \|\nabla^j F(\tilde{x}_t(j))\|^2 - \frac{D_{t+1}-D_t}{\eta_t} \nonumber\\
    & +\sum_{j=1}^m <x_t^j-x_*^j, \nabla^j F_t(x_t) -\nabla^j F_t(\tilde{x}_t(j))>.
\end{align}
\end{lemma}
\begin{align}
     & \text{{\bf Proof.}}\quad D_{t+1} - D_t \nonumber\\
    =& \frac{1}{2}\sum_{j=1}^{m} \big(\|x_{t}^j-\eta_t \nabla^j F(\tilde{x}_t(j))-x_*^j \|^2 - \|x_t^j-x_*^j\|^2\big)\nonumber\\
    =& \sum_{j=1}^{m} \big(\frac{1}{2}\|\eta_t \nabla^j F(\tilde{x}_t(j))\|^2 - \eta_t<x_t^j-x_*^j, \nabla^j F(\tilde{x}_t(j))>\big)\nonumber\\
    =& \frac{1}{2}\eta_t^2 \sum_{j=1}^{m} \|\nabla^j F(\tilde{x}_t(j))\|^2 - \eta_t<x_t-x_*, \nabla F(x_t)>\nonumber\\
    & + \eta_t \sum_{j=1}^{m} \big <x_t^j-x_*^j, \nabla^j F(x_t) -\nabla^j F(\tilde{x}_t(j))>.
\end{align}
Dividing the above equation by $\eta_t$ and rearranging it, we can get the lemma.
\hfill$\square$\\

Another important fact for our analysis is 
\begin{align}
    \sum_{t=a}^b \frac{1}{\sqrt{t}} \le \int_{a-1}^{b}\frac{1}{\sqrt{t}}dt = 2(\sqrt{b} - \sqrt{a-1}).\label{eq:sum_inequal}
\end{align}

We now come to evaluate the regret $R$ up to iteration $T$. By the definition in \eqref{eq:def_of_R} and, we have 

\begin{align}
    R &= \frac{1}{T}\sum_{t} F_t(x_t) - F(x_*)
    = \frac{1}{T}\sum_{t=1}^T F_t(x_t) - \frac{1}{T}\sum_{t=1}^T F_t(x_*)\\
    &= \frac{1}{T}\sum_{t=1}^T \big(F_t(x_t) - F_t(x_*)\big)
    \le \frac{1}{T}\sum_{t=1}^T  <x_t-x_*, \nabla F_t(x_t)>\label{eq:R_transform:line:convex}
\end{align}
where \eqref{eq:R_transform:line:convex} follows from the convexity of the loss functions. Inserting the result from lemma~\ref{lamma:main}, we can get
\begin{align}
    T\cdot R \le& \sum_{t=1}^T\big(\frac{1}{2}\eta_t \sum_{j=1}^{m} \|\nabla^j F(\tilde{x}_t(j))\|^2 - \frac{D_{t+1}-D_t}{\eta_t} \nonumber\\
    & +\sum_{j=1}^m <x_t^j-x_*^j, \nabla^j F_t(x_t) -\nabla^j F_t(\tilde{x}_t(j))>\big) \nonumber\\
    =& \sum_{t=1}^T\frac{1}{2}\eta_t \sum_{j=1}^{m} \|\nabla^j F(\tilde{x}_t(j))\|^2 - \sum_{t=1}^T\frac{D_{t+1}-D_t}{\eta_t} \nonumber\\
    & +\sum_{t=1}^T\sum_{j=1}^m <x_t^j-x_*^j, \nabla^j F_t(x_t) -\nabla^j F_t(\tilde{x}_t(j))>.\label{eq:R_lemma_in:line:sum_up}
\end{align}
We look into the three terms of \eqref{eq:R_lemma_in:line:sum_up} and bound them. 

For the first term, we have
\begin{align}
    \sum_{t=1}^T\frac{1}{2}\eta_t \sum_{j=1}^{m} \|\nabla^j F(\tilde{x}_t(j))\|^2
    \le & \sum_{t=1}^T \frac{1}{2}\eta_t m G^2
    = \sum_{t=1}^T \frac{1}{2}\frac{\eta}{\sqrt{t}} m G^2
    \le  \eta m G^2 \sqrt{T}.\label{eq:first_bound:line:last}
\end{align}
For the second term, we have 
\begin{align}
    - \sum_{t=1}^T \frac{D_{t+1}-D_t}{\eta_t}\ =& \frac{D_1}{\eta_1} - \frac{D_{t+1}}{\eta_t} + \sum_{t=2}^T D_t\left(\frac{1}{\eta_t} - \frac{1}{\eta_{t-1}}\right)\\
    \le&  \frac{D^2}{\eta} - 0 + \sum_{t=2}^T \frac{D^2}{\eta}\left(\sqrt{t} - \sqrt{t-1}\right)
    = \frac{D^2\sqrt{T}}{\eta}.\label{eq:second_bound:line:last}
\end{align}
Finally we come to the third term. We have

\begin{align}
    & \sum_{t=1}^T\sum_{j=1}^m <x_t^j-x_*^j, \nabla^j F_t(x_t) -\nabla^j F_t(\tilde{x}_t(j))>\nonumber\\
    \le & \sum_{t=1}^T\sum_{j=1}^m \|x_t^j-x_*^j\|\cdot\| \nabla^j F_t(x_t) -\nabla^j F_t(\tilde{x}_t(j))\| \label{eq:third:tri}\\
    \le & \sum_{t=1}^T\sum_{j=1}^m \|x_t^j-x_*^j\|\cdot L_j\| x_t - \tilde{x}_t(j) \| \label{eq:third:asu}\\
    \le & \sum_{t=1}^T\sum_{j=1}^m L_j\|x_t^j-x_*^j\|\cdot \sum_{i=1}^m \| x_t^i -  x^i_{t - \tau^j(i)} \|.\label{eq:third_term_mid_0}
\end{align}
If $\tau^j(i)\ge 0$, we have
\begin{align}
     \| x_t^i -  x^i_{t - \tau^j(i)} \| & 
     =  \big\|\sum_{q=t- \tau^j(i)}^{t-1} (x_{q+1}^i - x_q^i) \big\| 
     \le \sum_{q=t- \tau^j(i)}^{t-1} \|\eta_q \nabla^i F_q(\tilde{x}_q(i)) \|\nonumber\\
     & \le \sum_{q=t- \tau^j(i)}^{t-1} \eta_q G.\label{eq:third_term_mid_1}
\end{align}
If $\tau^j(i) < 0$, by similar technique, we can also get \eqref{eq:third_term_mid_1}. Inserting \eqref{eq:third_term_mid_1} into \eqref{eq:third_term_mid_0}, we have
\begin{align}
    & \sum_{t=1}^T\sum_{j=1}^m <x_t^j-x_*^j, \nabla^j F_t(x_t) -\nabla^j F_t(\tilde{x}_t(j))>\nonumber\\
    \le & Gm \sum_{t=1}^T\sum_{j=1}^m L_j\|x_t^j-x_*^j\|\cdot \sum_{q=t- \tau}^{t-1} \eta_q  \\
    \le & GmL_{\text{max}} \sum_{t=1}^T \sum_{q=t- \tau}^{t-1} \eta_q \sum_{j=1}^m \|x_t^j-x_*^j\|  \\
    \le & Gm^{\frac{3}{2}}L_{\text{max}} \sum_{t=1}^T \sum_{q=t- \tau}^{t-1} \eta_q \|x_t-x_*\| \label{eq:third:cauchy} \\
    \le & GDm^{\frac{3}{2}}L_{\text{max}} \sum_{t=1}^T \sum_{q=t- \tau}^{t-1} \eta_q.\label{eq:third_term_first}
\end{align}
\eqref{eq:third:tri} is from triangle inequality. \eqref{eq:third:asu} comes from the Assumption~\ref{theorem:assumption}'s blockwise Lipschitz continuity. \eqref{eq:third:cauchy} comes from the fact
\begin{align}
    \frac{1}{m}\sum_{j=1}^m \|x_t^j - x_*^j \| \le \frac{1}{\sqrt{m}}\sqrt{\sum_{j=1}^m \|x_t^j - x_*^j \|^2}
    = \frac{1}{\sqrt{m}}\|x_t - x_*\|.
\end{align}
For the last parts of \eqref{eq:third_term_first}, we have
\begin{align}
    \sum_{t=1}^T \sum_{q=t- \tau}^{t-1} \eta_q \le & \sum_{t=1}^\tau \eta_1 t + \sum_{t=\tau+1}^T\sum_{q=t-\tau}^{t-1} \eta_q
    \le  \frac{\eta \tau(\tau+1)}{2} + \sum_{t=\tau+1}^T \frac{\tau\eta}{\sqrt{t-\tau}}\\
    \le & \frac{\eta \tau(\tau+1)}{2} + 2\tau\eta\sqrt{T-\tau}\label{eq:third:second_sum}\\
    = & \frac{\eta\tau}{2}(\tau+1+4\sqrt{T}),\label{eq:eta_double_sum}
\end{align}
where \eqref{eq:third:second_sum} is from the fact \eqref{eq:sum_inequal}.
Combining \eqref{eq:third_term_first} and \eqref{eq:eta_double_sum}, we get 
\begin{align}
    & \sum_{t=1}^T\sum_{j=1}^m <x_t^j-x_*^j, \nabla^j F_t(x_t) -\nabla^j F_t(\tilde{x}_t(j))> \nonumber\\
    \le & \frac{1}{2}GDm^{\frac{3}{2}}L_{\text{max}}\eta\tau(\tau+1+4\sqrt{T}) \label{eq:third_bound:line:last}
\end{align}
Combining \eqref{eq:R_lemma_in:line:sum_up}, \eqref{eq:first_bound:line:last}, \eqref{eq:second_bound:line:last} and \eqref{eq:third_bound:line:last}, and dividing by $T$, we have
\begin{align}
    R \le & \frac{\eta m G^2}{\sqrt{T}} + \frac{D^2}{\eta\sqrt{T}} + \frac{1}{2}GDm^{\frac{3}{2}}L_{\text{max}}\eta\tau\frac{1}{\sqrt{T}}(\frac{\tau+1}{\sqrt{T}}+4)
    = O(\frac{1}{\sqrt{T}}).\nonumber
\end{align}
\hfill$\blacksquare$

\end{document}